\documentclass[a4paper]{jpconf}
\usepackage[utf8]{inputenc}
\usepackage[english]{babel}
\usepackage{amsmath,amssymb}
\usepackage{textcomp,gensymb}
\usepackage{graphicx}
\usepackage{cite}

\begin{document}
\title{Quantum disordered ground state\\ for the frustrated square lattice}

\author{A Kalz, A Honecker, S Fuchs and T Pruschke}

\address{Institut für Theoretische Physik,  Georg-August-Universität Göttingen, \\ Friedrich-Hund-Platz 1, 37077 Göttingen, Germany}

\ead{kalz@theorie.physik.uni-goettingen.de}

\begin{abstract}
We present Quantum Monte-Carlo simulations of an exchange-anisotropic spin-$1/2$ Heisenberg model on a square lattice with nearest and next-nearest neighbor interactions. The ground state phase diagram shows two classical magnetically ordered phases for dominating antiferromagnetic $S^z$-interactions and for large quantum fluctuations a ferromagnetic order in the $x$-$y$ plane. In between a finite region is detected where neither classical nor quantum mechanical order, e.g. long-ranged dimer correlations, are found. 
\end{abstract}

\section{Introduction \label{s:intro}}
The investigation of frustrated spin models has become a rather active field in the past decades \cite{B:richter04, B:diep, B:mila11}. In particular the appearance of quantum critical behavior and emergence of new quantum phases in these systems has connected this field of low temperature physics directly with the research on high-$T_C$ superconductors and topological insulators \cite[and references therein]{P:ihle90, P:manousakis91, P:nori92, P:bergman08, P:guo09}. The frustration induced suppression of classical order especially for systems in two dimensions and the corresponding enhancement of quantum fluctuations yield an interesting competition between ordered ground states -- like staggered dimer phases -- and quantum disordered states as the spin liquid state \cite{P:bartosch05, P:sachdev08, P:kalz11}.

The analysis of such frustrated models with Quantum Monte-Carlo (QMC) simulations is often complicated due to the sign problem which limits the ability of the algorithm for competing quantum fluctuations. This is one reason why in the present work we choose the quantum fluctuations to be non-frustrating (i.e., the $S^{xy}$-interactions are ferromagnetic). In addition such a spin model can be mapped exactly onto a model of hard-core bosons with repulsive interactions \cite{P:batrouni00, P:batrouni01}. These bosonic models can be analyzed in optical lattices \cite{P:lewenstein07} and also frustrated systems have been realized recently \cite{P:jo11}.

In the following section we will introduce the model and describe the expected ground states before explaining our QMC results in Sec.~\ref{s:results}. We conclude with a short discussion in Sec.~\ref{s:dis}.
 
\section{Model \label{s:model}}
The model is given by summing over all interactions on nearest neighbor (NN) and next-nearest neighbor (NNN) bonds between spin-$1/2$ operators on a $L\times L$ square lattice:
\begin{align}
H&= \sum_{\text{NN}} t_1 (S^+_iS^-_j + \text{H.c.})+V_1~S^z_iS^z_j
+ \sum_{\text{NNN}} t_2 (S^+_iS^-_j + \text{H.c.})+V_2~S^z_iS^z_j\,. \label{eq:model}
\end{align}
The $t_k<0$ couple the $S^{xy}$-direction ferromagnetically and the $V_k>0$ are chosen antiferromagnetic and, hence, introduce frustration in the model. The nearest neighbor $S^z$-coupling favors a Néel state which has the lowest energy in the classical Ising case ($t_k=0$) for values $V_2 < V_1/2$ and competes with a collinear configuration, i.e., parallel spins in one direction and antiparallel spins in the other direction, at the critical point $V_2 = V_1 /2$. Classically the ground state is highly degenerate at this point \cite{P:kalz08, P:kalz09}. For large quantum fluctuations ($|t_k| \approx 1/2$) a ferromagnetic alignment of all spins in the $x$-$y$ plane is expected. This corresponds to a superfluid state of the above mentioned hard-core bosons \cite{P:bloch31, P:matsubara56}.

In the vicinity of the critical point $V_2 = V_1 / 2$ and for non-zero fluctuations the emergence of quantum ordered phases is possible since quantum fluctuations play a crucial role in low dimensional systems. This was discussed for similar models, e.g.\,in \cite{P:bartosch05}. These phases consist of quantum mechanically entangled states coupling two or more spins on neighboring sites. Examples are staggered or columnar configurations of dimers on nearest-neighbor bonds. However, for the present model these dimers are not given by the spin singlet state of two spins but due to the exchange anisotropy by the $S^z = 0$ triplet state.

\begin{figure}[t!]
\begin{minipage}[][0.6\linewidth][t]{0.48\linewidth}
\includegraphics[width=\linewidth]{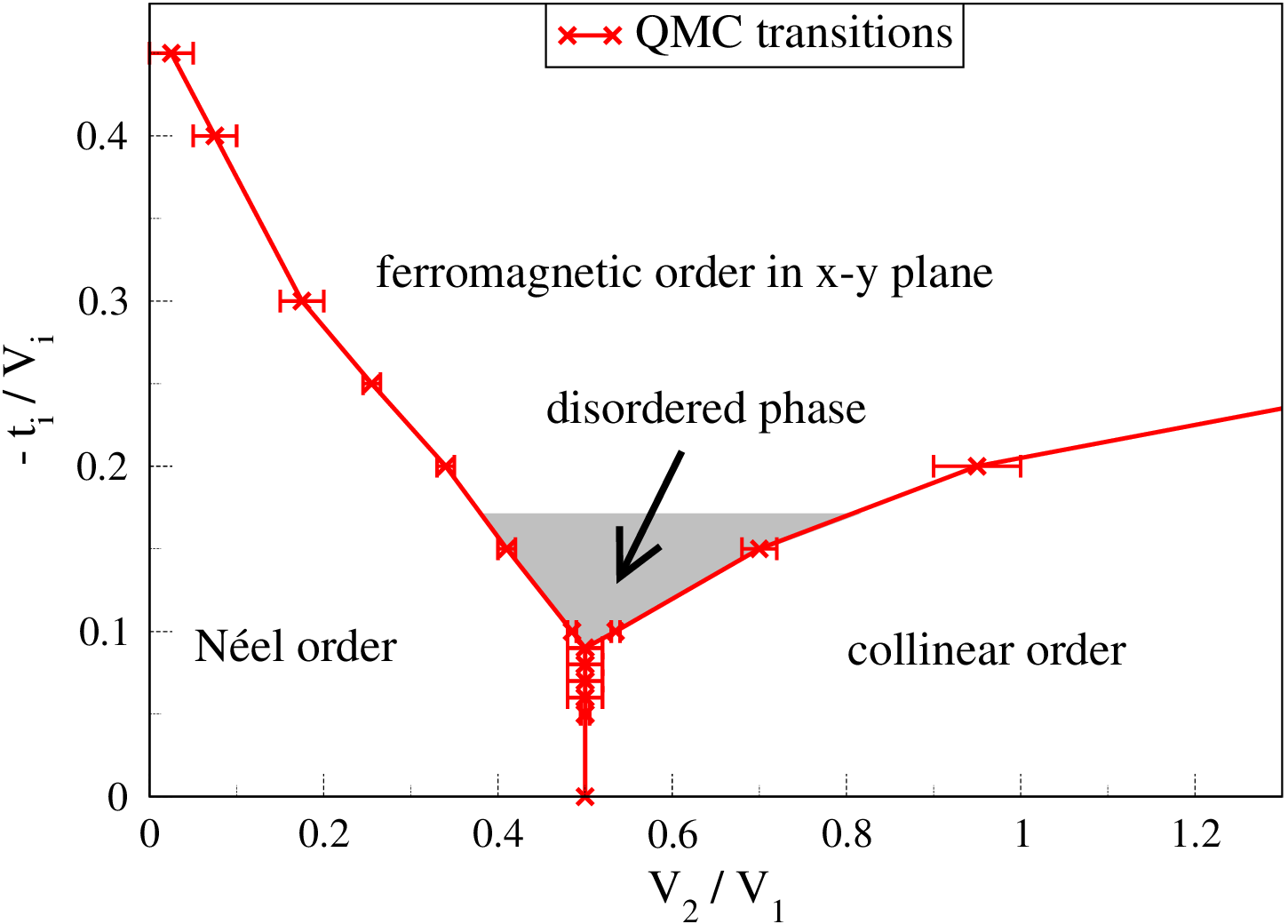}
\caption{\label{f:phase}Ground state phase diagram of the anisotropic Heisenberg model on the frustrated square lattice (data from our previous work \cite{P:kalz11}). For small fluctuations $|t_k|$ the classical antiferromagnetic phases survive and for large values a ferromagnetic order in the $x$-$y$ plane is stabilized. For the intermediate region we find no finite signal of any order parameter.}
\end{minipage}
\hspace*{0.02\linewidth}
\begin{minipage}[][0.6\linewidth][t]{0.48\linewidth}
\includegraphics[width=\linewidth]{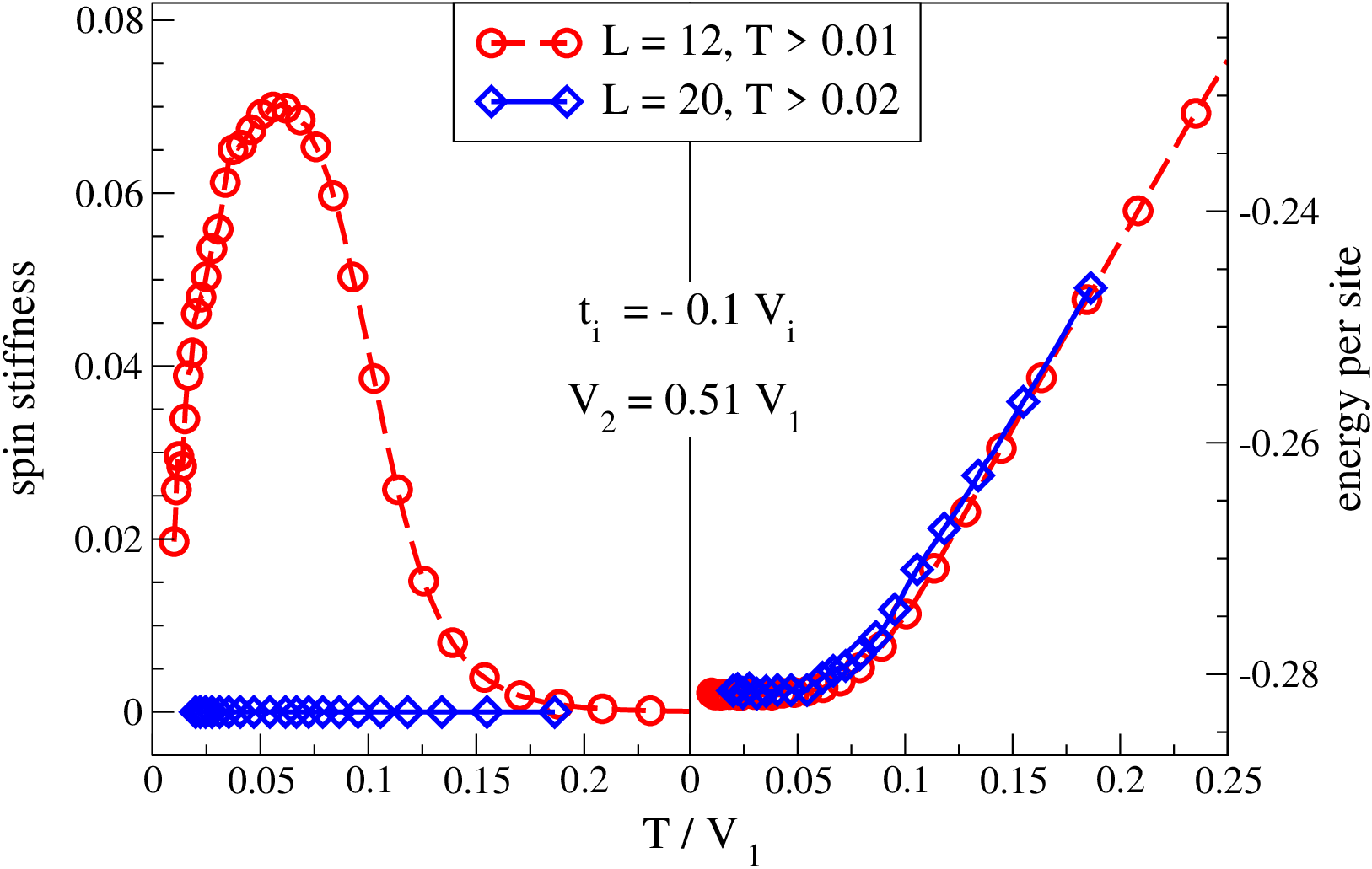}
\caption{\label{f:stiff} \textbf{Left: }The spin stiffness inside the disordered region at $t_k = -0.1~V_k$ and $V_2 = 0.51~V_1$ only shows a small signal for an intermediate temperature on the $12\times 12$ lattice. For a larger lattice no finite signal is found. \textbf{Right: }The energies are saturated to their ground state value for $T < 0.05~V_1$.}
\end{minipage}
\end{figure}
\section{Results \label{s:results}}
The phase diagram of the model \eqref{eq:model} is given in Fig.~\ref{f:phase} in terms of the strength of fluctuations $-t_1/V_1=-t_2/V_2$ and frustration $V_2/V_1$. The phase transition lines were derived in Ref.~\cite{P:kalz11} from finite-temperature QMC simulations using an implementation of the \emph{Stochastic Series Expansion} \cite{P:sandvik91, P:sandvik02} which is part of the ALPS project \cite{P:alet05, P:ALPS05, P:ALPS07}. Furthermore, an exchange Monte-Carlo update \cite{P:hukushima96, P:hansmann97, P:katzgraber06, P:melko07} was used to overcome thermalization problems in the vicinity of the critical point $V_2=V_1/2$. 

To detect the different magnetic phases we calculated the structure factor which is given by the Fourier transformed spin-spin correlation function for the wave vectors $\vec q=(\pi,\pi)$ (Néel order) and $\vec q = (\pi,0),(0,\pi)$ (collinear order). 
The ferromagnetic phase -- corresponding to superfluid order in an analogue bosonic model -- can be detected by calculating the spin stiffness (or superfluid density). In the QMC simulations the estimator for this observable is connected with the winding number \cite{P:pollock87}.

As expected the classical antiferromagnetic states survive for small $|t_k|$ and a direct transition between both states is found. However, above the critical point we find for $|t_k| \geq 0.1~V_k$ an opening of a region without classical order in the phase diagram. We performed careful calculations of the spin stiffness in this region and found it to be suppressed for lattice sizes $L \geq 16$ and low temperatures $T < 0.05~V_1$ (Fig.~\ref{f:stiff}). Only for larger values of $|t_k|>0.175~V_k$ the ferromagnetic order parameter shows a finite and size-independent signal \cite{P:kalz11}.

\begin{figure}[t!]
\begin{minipage}[][0.6\linewidth][t]{0.48\linewidth}
\includegraphics[height=\linewidth,angle=270]{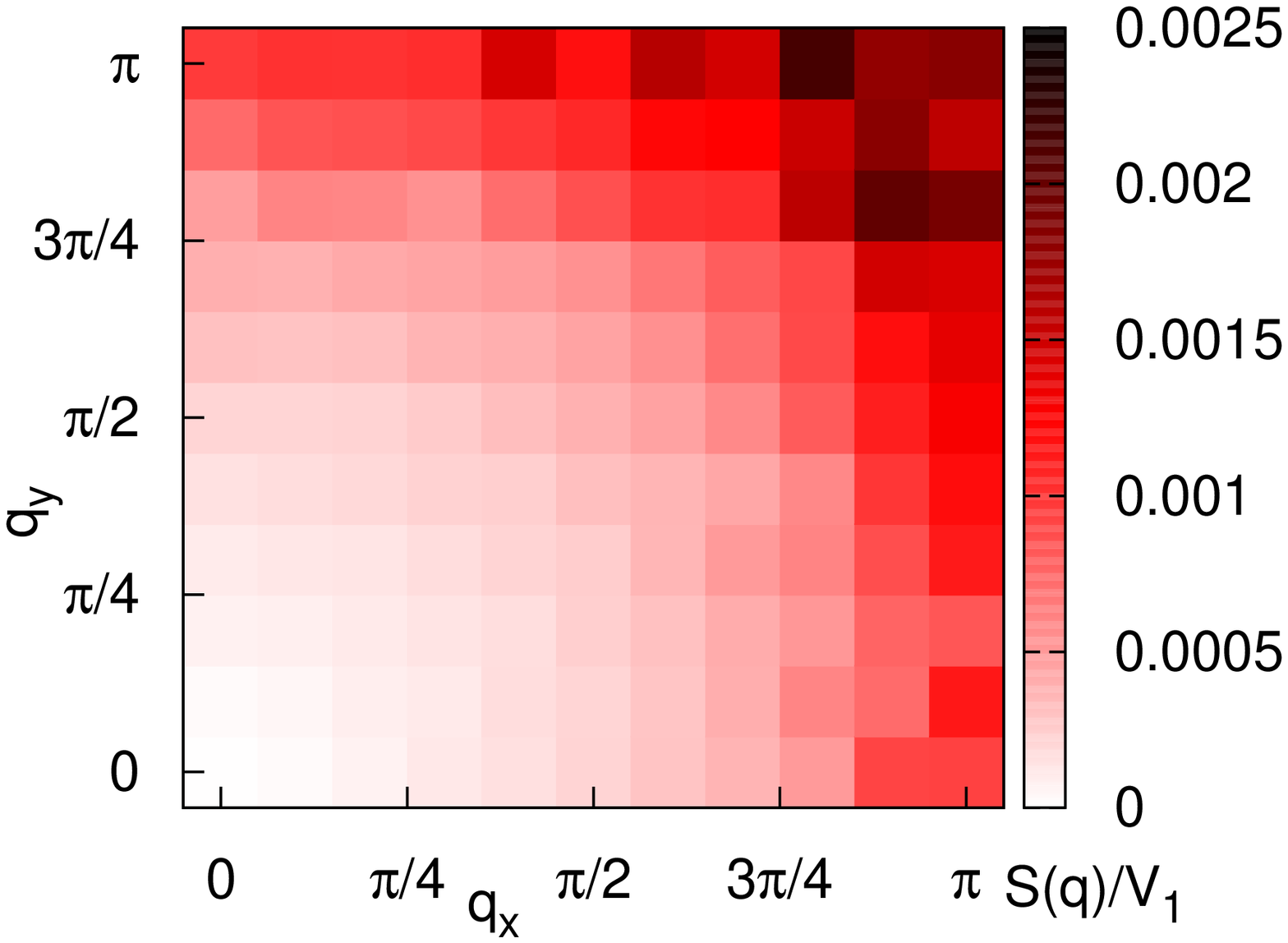}
\caption{\label{f:structure} For $t_k = -0.1~V_k$ and $V_2=0.51~V_1$ the complete structure factor of a $20 \times 20$ lattice is shown and only a vanishingly small signal at $\vec q = (\pi, \pi)$ (Néel type) is identified -- probably due to finite-size effects and the finite temperature ($T=0.02~V_1$).}
\end{minipage}
\hspace*{0.02\linewidth}
\begin{minipage}[][0.62\linewidth][t]{0.48\linewidth}
\includegraphics[width=\linewidth]{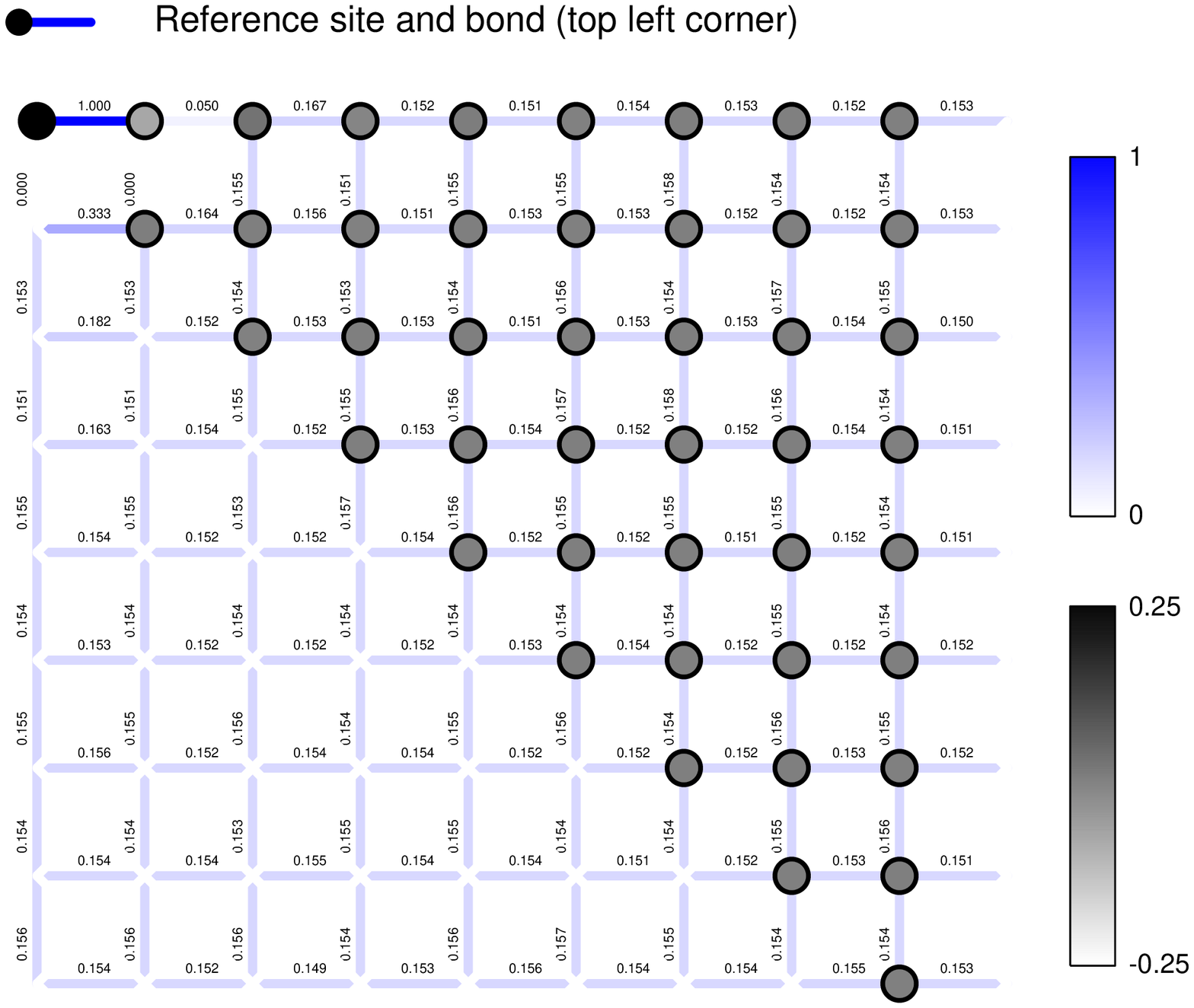}
\caption{\label{f:dimer} The correlation of spins and dimers with respect to the top left site and adjacent bond are shown in a color coded lattice for $t_k = - 0.12~V_k$ and $V_2= 0.5~V_1$ at $T=0.01~V_1$. No sign of long-ranged order is visible neither in the spin-spin correlation function nor in any configuration of parallel dimers or plaquettes.}
\end{minipage}
\end{figure}

For a set of parameters in the shaded region of the phase diagram in Fig.~\ref{f:phase} we calculated further observables. In Fig.~\ref{f:structure} we show as an example the complete structure factor of a $20 \times 20$ lattice in a color-coded map. Apart from a small enhancement of the values for $\vec q \approx (\pi, \pi)$ which occurs due to finite-size effects and thermal fluctuations ($T=0.02~V_1$) we find no evidence for any magnetic order in the $S^z$-components of commensurate or incommensurate type.

Since no classical order is found, the emergence of quantum ordered phases as the columnar or staggered configuration of dimers was also tested. Therefore, we implemented an estimator for the correlation of four spins residing on two different nearest-neighbor bonds \cite{P:sandvik92}. The result of such a calculation is shown in Fig.~\ref{f:dimer} for a $16\times 16$ lattice and parameters inside the unidentified region. The correlation of parallel and perpendicular dimers on nearest-neighbor bonds is plotted in a color-coded lattice which reflects the relative distance between the top left dimer and a second dimer. The strength of the correlation is normalized to the value of the self-correlation of the reference bond. We also include the results of spin-spin correlation functions on the lattice sites. We cannot identify any long-ranged order in the dimer or spin correlations and, hence, conclude that for a small region of the phase diagram the system is in a quantum disordered ground state.

\section{Discussion \label{s:dis}}
We performed QMC simulations for an anisotropic Heisenberg model on a square lattice with competing interactions. The phase diagram was derived and a finite region without classical order was identified. An additional analysis of higher order correlation functions also ruled out the appearance of dimer ordered phases. In our recent work \cite{P:kalz11} we also presented exact diagonalization results for a set of parameters inside the disordered region and found a finite gap in the energy spectrum which is compatible with the rapid decay of the spin-spin correlation functions shown in Fig.~\ref{f:dimer}. We conclude from our simulations that we have found a quantum mechanically disordered ground state with a finite gap which can be referred to as a gapped spin liquid.

Investigations of a similar model on the honeycomb lattice with additional third-nearest neighbor interactions are in progress. This model is known to exhibit emergent quantum phases for the exchange-isotropic case \cite{P:fouet01, P:cabra10, P:albuquerque11}.

\section*{Acknowledgments}
We acknowledge financial support by the Deutsche Forschungsgemeinschaft under grant No. HO 2325/4-2 and through SFB 602.

\section*{References}
\bibliography{Literatur_sces.bib}

\providecommand{\newblock}{}
\begin{thebibliography}{10}
\expandafter\ifx\csname url\endcsname\relax
  \def\url#1{{\tt #1}}\fi
\expandafter\ifx\csname urlprefix\endcsname\relax\def\urlprefix{URL }\fi
\providecommand{\eprint}[2][]{\url{#2}}

\bibitem{B:richter04}
Richter J, Schulenburg J and Honecker A 2004 {\em Quantum magnetism\/} ({\em
  Lecture Notes in Physics\/} vol 645) ed Schollw\"ock U, Richter J, Farnell
  D~J~J and Bishop R~F (Springer) chap~2, p~85

\bibitem{B:diep}
Diep H~T (ed) 2005 {\em Frustrated Spin Systems\/} (World Scientific)

\bibitem{B:mila11}
Lacroix C, Mendels P and Mila F (eds) 2011 {\em Introduction to Frustrated
  Magnetism\/} 1st ed ({\em Springer Series in Solid-State Sciences\/} vol 164)
  (Springer)

\bibitem{P:ihle90}
Ihle D and Kasner M 1990 {\em Phys. Rev. B\/} {\bf 42} 4760--4762

\bibitem{P:manousakis91}
Manousakis E 1991 {\em Rev. Mod. Phys.\/} {\bf 63} 1--62

\bibitem{P:nori92}
Nori F, Gagliano E and Bacci S 1992 {\em Phys. Rev. Lett.\/} {\bf 68} 240--243

\bibitem{P:bergman08}
Bergman D~L, Wu C and Balents L 2008 {\em Phys. Rev. B\/} {\bf 78} 125104

\bibitem{P:guo09}
Guo H~M and Franz M 2009 {\em Phys. Rev. B\/} {\bf 80} 113102

\bibitem{P:bartosch05}
Balents L, Bartosch L, Burkov A, Sachdev S and Sengupta K 2005 {\em Phys. Rev.
  B\/} {\bf 71} 144508

\bibitem{P:sachdev08}
Sachdev S 2008 {\em Nature Physics\/} {\bf 4} 173--185

\bibitem{P:kalz11}
Kalz A, Honecker A, Fuchs S and Pruschke T 2011 {\em Phys. Rev. B\/} {\bf 83}
  174519

\bibitem{P:batrouni00}
Batrouni G~G and Scalettar R~T 2000 {\em Phys. Rev. Lett.\/} {\bf 84}
  1599--1602

\bibitem{P:batrouni01}
H\'ebert F, Batrouni G~G, Scalettar R~T, Schmid G, Troyer M and Dorneich A 2001
  {\em Phys. Rev. B\/} {\bf 65} 014513

\bibitem{P:lewenstein07}
Lewenstein M, Sanpera A, Ahufinger V, Damski B, Sen A and Sen U 2007 {\em
  Advances in Physics\/} {\bf 56} 243--379

\bibitem{P:jo11}
Jo G, Guzman J, Thomas C, Hosur P, Vishwanath A and Stamper-Kurn D 2011 {\em
  cond-mat:1109.1591\/}

\bibitem{P:kalz08}
Kalz A, Honecker A, Fuchs S and Pruschke T 2008 {\em Eur. Phys. J. B\/} {\bf
  65} 533--537

\bibitem{P:kalz09}
Kalz A, Honecker A, Fuchs S and Pruschke T 2009 {\em J. Phys.: Conf. Ser.\/}
  {\bf 145} 012051

\bibitem{P:bloch31}
Bloch F 1932 {\em Z. Phys.\/} {\bf 74} 295--335

\bibitem{P:matsubara56}
Matsubara T and Matsuda H 1956 {\em Prog. Theor. Phys.\/} {\bf 16} 569--582

\bibitem{P:sandvik91}
Sandvik A~W and Kurkij\"arvi J 1991 {\em Phys. Rev. B\/} {\bf 43} 5950--5961

\bibitem{P:sandvik02}
Sylju\aa{}sen O~F and Sandvik A~W 2002 {\em Phys. Rev. E\/} {\bf 66} 046701

\bibitem{P:alet05}
Alet F, Wessel S and Troyer M 2005 {\em Phys. Rev. E\/} {\bf 71} 036706

\bibitem{P:ALPS05}
Alet F \textit{et al} 2005 {\em J. Phys. Soc. Jpn.\/} {\bf 74S} 30

\bibitem{P:ALPS07}
Albuquerque A~F \textit{et al} 2007 {\em J. Magn. Magn. Mat.\/} {\bf 310} 1187

\bibitem{P:hukushima96}
Hukushima K and Nemoto K 1996 {\em J. Phys. Soc. Jpn.\/} {\bf 65} 1604--1608

\bibitem{P:hansmann97}
Hansmann U~H~E 1997 {\em Chem. Phys. Lett.\/} {\bf 281} 140--150

\bibitem{P:katzgraber06}
Katzgraber H~G, Trebst S, Huse D~A and Troyer M 2006 {\em J. Stat. Mech.:
  Theory and Experiment\/}  P03018

\bibitem{P:melko07}
Melko R~G 2007 {\em J. Phys.: Cond. Matt.\/} {\bf 19} 145203

\bibitem{P:pollock87}
Pollock E~L and Ceperley D~M 1987 {\em Phys. Rev. B\/} {\bf 36} 8343--8352

\bibitem{P:sandvik92}
Sandvik A~W 1992 {\em J. Phys. A\/} {\bf 25} 3667--3682

\bibitem{P:fouet01}
Fouet J, Sindzingre P and Lhuillier C 2001 {\em Eur. Phys. J. B\/} {\bf 20}
  241--254

\bibitem{P:cabra10}
Cabra D, Lamas C and Rosales H 2010 {\em Phys. Rev. B\/} {\bf 83} 094506

\bibitem{P:albuquerque11}
Albuquerque A~F, Schwandt D, Het\'enyi B, Capponi S, Mambrini M and L\"auchli
  A~M 2011 {\em Phys. Rev. B\/} {\bf 84} 024406

\end{thebibliography}
\bibliographystyle{iopart-num.bst}

\end{document}